\documentstyle[draft]{article}

\catcode`\@=11

\def\journal{\topmargin .3in \oddsidemargin .5in \headheight 0pt
  \headsep 0pt \textwidth 5.625in \textheight 8.25in \marginparwidth 1.5in
  \parindent 2em \parskip .5ex plus .1ex \jot = 1.5ex}
\def\preprint{\twocolumn\sloppy\flushbottom\parindent 2em \leftmargini 2em
  \leftmarginv .5em \leftmarginvi .5em \oddsidemargin -.5in
  \evensidemargin -.5in	\columnsep .4in	\footheight 0pt	\textwidth 10in
  \topmargin  -.4in \headheight 12pt \topskip .4in \textheight 7.1in
  \footskip 0pt
  \def\@oddhead{\thepage\hfil\addtocounter{page}{1}\thepage}
  \let\@evenhead\@oddhead \def\@oddfoot{} \def\@evenfoot{} }
\def\titlepage{\@restonecolfalse\if@twocolumn\@restonecoltrue\onecolumn
     \else \newpage \fi \thispagestyle{empty}\c@page\z@
	\def\thefootnote{\fnsymbol{footnote}} }
\def\endtitlepage{\if@restonecol\twocolumn \else \newpage \fi
	\def\thefootnote{\arabic{footnote}}
	\setcounter{footnote}{0}}  
\def\section{\@startsection {section}{1}{0pt}{-3.5ex plus -1ex minus
   -.2ex}{2.3ex plus .2ex}{\raggedright\large\bf}}
\def\numberbysection{\@addtoreset{equation}{section}
	\def\theequation{\thesection.\arabic{equation}}}
\catcode`\@=12

\journal
\numberbysection  
\preprint 

\def\frac#1#2{{\textstyle {#1 \over #2}}}
\def\Eq{\begin{equation}}	\def\End{\end{equation}}
\def\Eqa{\begin{eqnarray}}	\def\Enda{\end{eqnarray}}
\def\Endl#1{\label{#1} \End}	\def\Endla#1{\label{#1} \Enda}
\def\puteq#1{eq.~(\ref{#1})}	\def\d{{\rm d}}
\def\M{{\cal M}}		\def\bp{\bar{\phi}}
\def\p{{\bf p}}			\def\gev{{\rm\,GeV}}
\def\O{{\cal O}}		\def\Im{{\rm Im}}
\def\om{\omega_n}		\def\Tr{{\rm Tr~}}
\def\mw{m_W^2}			\def\mc{m_\chi^2}
\def\bm{\tilde m}
\def\sik{T \sum_n \int {\d^3 {\bf k} \over (2\pi)^3}}
\def\sip{T \sum_m \int {\d^3 {\bf p} \over (2\pi)^3}}

\begin{document}
\begin{titlepage}
\begin{center}
June, 1992			\hfill       CALT-68-1795\\
				\hfill       HUTP-92-A027\\
				\hfill       EFI-92-22\\
\vskip .5in {\large\bf
Corrections to the Electroweak Effective Action at Finite Temperature%
}\footnote{This work was supported by the Director, Office of Energy
Research, Office of High Energy and Nuclear Physics, Division of High
Energy Physics of the U.S. Department of Energy under contracts
DEAC-03-81ER40050 and DEFG-02-90ER40560.}\\
\vskip.3 in
{
  {\bf C. Glenn Boyd}\footnote{Email: \tt boyd@rabi.uchicago.edu}\\
  \vskip.2 cm {\it Enrico Fermi Institute, 5640 Ellis Ave., Chicago IL 60637}
\vskip.3 cm
  {\bf David E. Brahm}\footnote{Email: \tt brahm@theory3.caltech.edu}\\
  \vskip.2 cm {\it Caltech 452-48, Pasadena CA 91125}
\vskip.3 cm and \vskip.3 cm
  {\bf Stephen D.H. Hsu}\footnote{Junior Fellow, Harvard Society of
     Fellows. Email: \tt hsu@hsunext.harvard.edu }
  \vskip.2 cm {\it Lyman Laboratory of Physics, Harvard University,
     Cambridge MA 02138}\\ }
\end{center}

\vskip.5 in
\begin{abstract}
We calculate contributions to the finite temperature effective action
for the electroweak phase transition (EWPT) at $\O(g^4)$, {\it i.e.} at
second order in $(g^2 T/\M)$ and all orders in $(g^2 T^2/\M^2)$.  This
requires plasma-mass corrections in the calculation of the effective
potential, inclusion of the ``lollipop'' diagram, and an estimate of
derivative corrections.  We find the EWPT remains too weakly first-order to
drive baryogenesis.  We calculate some one loop kinetic energy corrections
using both functional and diagrammatic methods; these may be important for
saddlepoint configurations such as the bounce or sphaleron.
\end{abstract}
\end{titlepage}

\renewcommand{\thepage}{\arabic{page}}
\setcounter{page}{1}


\section{Introduction}

Recent work \cite{dine,baryo} suggests the baryon asymmetry may have been
generated at the electroweak phase transition (EWPT).  This would require
the transition be first-order \cite{oldba}, with the resulting Higgs vev
large (roughly, $\phi_+(T_b)/T_b > 1.4$).  Several authors \cite{dine,AH},
using the 1-loop finite temperature effective potential
\cite{dj,wein,linde}, have concluded that these requirements may be met in
the Standard Model (with augmented CP violation) for a sufficiently light
Higgs, say $M_h < 55\gev$ (now just below experimental limits \cite{LEP}).

Since the transition is weakly first-order, infrared divergences \cite{spt}
from the resultant nearly massless scalar and gauge boson modes make
higher-loop graphs important.  The inclusion of plasma masses
\cite{kap,fend} accounts for the most important corrections, $\O(g^2
T^2/\M^2)$, while all other higher-loop corrections are $\O(g^2 T/\M)$
\cite{us}.  Several authors have examined plasma mass corrections in the
gauge sector \cite{us,shap,carr,comm}; while we will refine these
calculations, the basic result holds that electric modes decouple
\cite{comm}, reducing the cubic term in $V$ by $1/3$.  For the Higgs sector
Carrington \cite{carr} computed the leading plasma
masses [$\O(g^2 T^2)$ and $\O(\lambda T^2)$], and Brahm and Hsu \cite{us}
worked to higher order in $g$.  Unfortunately, the vacuum-to-vacuum method
we used overcounts some contributions, as pointed out by the authors of
ref.~\cite{comm} and by Boyd; in this paper we re-examine the Higgs sector
using tadpole graphs.

One may question the validity of inserting zero-momentum plasma masses into
our diagrams \cite{arnold}; we estimate the error involved using the
derivative expansion of the effective action \cite{chan}, and also by
direct calculation of a two point graph.  We note derivative terms can be
important for determining critical bubbles or (B+L)-violating sphaleron
solutions \cite{sphal}, and resolving questions of gauge-invariance.

\section{The 1-loop effective potential}

The effective action $\Gamma[\bp]$ is the double Legendre transform of the
generator of 1PI truncated
Green's functions, and is the Legendre transform of $W[J]$:
  \Eq \Gamma [\bp] = W[J] - \int \d^4 x J(x) \bp(x), \Endl{Legend}
where $\bp$ is the expectation value of the field operator $\hat{\phi}$ in
the presence of source $J$.  The quantum theory described by $W[J]$ is
equivalent to a classical (tree-level) theory described by $\Gamma[\bp]$,
which can be expanded in derivatives,
  \Eq \Gamma[\bp] = \int d^4 x [ - V(\bp) + A
    (\partial_{\mu} \bp)^2 + \cdots ]. \Endl{Gamma}
On the restricted Hilbert space of states localized in $\phi$, $V$ is the
usual effective potential \cite{alex,ww}.

At finite (non-zero) temperature, $V$ can
be identified with the free-energy density in the convex region, and can
be calculated by imposing periodic (antiperiodic) boundary conditions on
bosonic (fermionic) fields in Euclidean time: $k_4^2 \to \omega_n^2 = (2\pi
n T)^2$ and $\int \frac{d^4 k}{(2 \pi)^4} \to T \sum_n \int\frac{d^3
k}{(2\pi)^3}$, where $n$ is an integer (half-integer) for bosons
(fermions) \cite{dj,wein,sher,epot}. $V$ calculated to 1-loop can be
written as the sum of tree-level, $T\!=\!0$, and finite-$T$ contributions:
  \Eq V = V_0 + V_1 + V_T, \qquad V_0(\phi) = {\lambda\over4} (\phi^2 -
    v^2)^2 \Endl{pieces}
where $\lambda = M_h^2/2v^2$. At any order, a useful approximate
parameterization \cite{AH,comm} is
   \Eq V = D(T^2 - T_0^2)\bp^2 -E T \bp^3 + { \lambda_T \over 4}\bp^4
       \Endl{param}
where $T_0$ is the temperature at which $V''(\bp=0)$ vanishes. This
parameterization can be used to estimate quantities such as $T_c$, the
temperature at which there are two degenerate minima, and $\bp_+$, the
position of the non-symmetric degenerate minimum.

We add counterterms \cite{AH} to $V_1$ to maintain $V'(v)=0$ and $V''(v) =
M_h^2 - \Sigma(M_h^2) + \Sigma(0)$ at $T\!=\!0$.  The latter relation
arises because $V$ is calculated at vanishing external momentum while the
physical Higgs mass is defined on-shell ($p^2=M_h^2$) \cite{aitch}; it is
accounted for by ``running'' $\lambda$ down to zero momentum:
  \Eq V_1 = {\Delta\lambda\over4} (\bp^2 - v^2)^2 +
    \sum_j {\pm g_j\over64\pi^2} \left\{ m_j^4 \ln\left[ m_j^2 \over M_j^2
    \right] - \frac32 m_j^4 + 2 m_j^2 M_j^2 \right\} \Endl{V1}
where
  \Eq \Delta\lambda = {-1\over 2v^2} \sum_j [\Sigma_j(M_h^2)-\Sigma_j(0)]
    \equiv \sum_j {\mp g_j M_j^4 \over 128 \pi^2 v^4} \, f_j(M_h^2/M_j^2)
  \Endl{fjdef}
The sums are over all particles $j$ with $g_j$ degrees of freedom and mass
$m_j(\phi)$; we write $M_j = m_j(v)$.  The upper sign is for bosons, the
lower for fermions.  The Standard Model fields in Landau
gauge \cite{dine,AH,cw} are the Higgs ($g_h=1$), the Goldstone bosons
($g_\chi=3$), the top quark ($g_t=12$), and the gauge bosons ($g_W=6$,
$g_Z=3$), with tree-level masses
  \Eq m_h^2 \!=\! \lambda (3\phi^2-v^2), \quad m_\chi^2 \!=\! \lambda
     (\phi^2-v^2), \quad m_t \!=\! {M_t \phi \over v}, \quad m_W \!=\! {g
     \phi \over 2}, \quad m_Z \!=\! {G \phi \over 2} \End
where $G^2 = g^2 + g^{\prime2}$.

In Appendix A we give the running functions $f_j(r)$ defined in
\puteq{fjdef}.  Note the Goldstone bosons, massless in Landau gauge
for $\bp = v$,
contribute a logarithmic infinity to $\Sigma(0)$ and thus to $\Delta\lambda$,
which exactly cancels the infinity in the $\ln[ m_\chi^2/M_\chi^2 ]$ term.

The $T$-dependent part of the effective potential is
\cite{dj,wein,sher,epot}
  \Eq V_T = g_j {T^4 \over 2\pi^2}\; I_\pm(m_j/T), \qquad
    I_\pm(y) \equiv \pm \int_0^\infty \d x \> x^2
    \ln\left( 1 \mp e^{-\sqrt{x^2+y^2}} \right) \Endl{Idef}
Series expansions of $I_\pm$ can be found in Appendix B.  The $\phi^3$ term
arising from gauge boson loops is primarily responsible for the hump in the
potential \cite{dine,AH,kirz}.

The Higgs and Goldstone masses can make imaginary contributions to $V$.
For homogeneous field configurations, these represent the rate of decay to
inhomogeneous states \cite{ww}.  We speculate that for the critical bubble
they are canceled (at least in large part) by derivative corrections
\cite{flash}.  The critical bubble has only one negative eigenmode (the
``breathing'' mode), whose contribution to the imaginary part of the action
appears to be independent of the bubble radius $R$ in the thin-wall limit,
whereas the imaginary parts of $V$ would contribute $\sim R^2$ to the
action if uncancelled.  It is also very suggestive that the region in which
the integrand of \puteq{Idef} is complex ($x<|y|$, or $k<|m|$) arises from
Fourier modes of $\phi$ with wavelengths larger than the bubble wall
thickness \cite{flash,GAco}.  We eliminate these modes by taking the real
part of $V_1$ and changing the lower limit of integration in \puteq{Idef}\
to $\Im\{y\}$.

In the tadpole method \cite{kirz,lee}, $V'(\bp)$ is calculated from tadpole
graphs using Feynman rules in the shifted theory $\phi = \bp + \phi'$,
dropping linear terms in $\phi'$ which are canceled in the Legendre
transform relating $W[J]$ and $\Gamma[\bp]$ [\puteq{Legend}]
\cite{JIIM,sher}.  The 1-loop diagrams are shown in Fig.~1a, giving (with
$\bm=m$ for now)
  \Eqa V' &=& (\lambda+\Delta\lambda) \bp (\bp^2-v^2) \nonumber\\
    &+& \sum_j {\pm g_j\over32\pi^2}{\d m_j^2 \over \d\bp} \left\{ \bm_j^2
      \left( \ln\left[ \bm_j^2 \over M_j^2 \right] - 1 \right) + M_j^2
      \right\} \nonumber\\
    &+& \sum_j {g_j T^2\over24}{\d m_j^2 \over \d\bp} F_\pm\left( \bm \over T
      \right)  \Endla{vprime}
where $F_\pm(y) \equiv 6 I'_\pm(y)/(\pi^2 y)$.

\begin{figure}
\vskip4.0in
\caption{Loop Diagrams for $V$}
\end{figure}

\section{Higher order corrections}

Beyond one loop, the most important diagrams are daisies and
super-daisies, as well as one other, the lollipop (Fig.~1e).

Consider the W tadpole (first diagram of Fig.~1a), which contributes $\sim
(g T^2 M)$ to $V'$, where $M \sim g\bp$.  ``Daisies'' \cite{dj} (Fig.~1b,c)
are diagrams with loops on the main W loop; those in Fig.~1b contribute
$\sim (g T^2 M) (g^2 T^2/M^2)^n (g^2 T/M)$. If we ignore powers of
$T / \bp $, daisies are $\O(g^3)$.
``Super-daisies'' (Fig.~1d) contribute $\sim (g T^2 M) (g^2 T^2/M^2)^n
(g^2 T/M)^2$, and are $\O(g^4)$ corrections.
While many other diagrams exist which cannot be classified as daisies or
super-daisies (Fig.~1e,f), only the ``lollipop'' (Fig.~1e) is $\O(g^4)$.
In the region of interest $(g^2 T^2/M^2) \approx 1$ and $(g^2
T/M) < 1$ \cite{wein}, so we make a consistent approximation by considering
super-daisy diagrams and the lollipop. In the Higgs sector the same
categorization holds by replacing $g^2 \to \lambda$.

We emphasize that the convergence properties of the loop expansion depend
on the choice of $\bp$ as well as T.  For a given temperature $T$, the
expansion may fail [$(g^2 T/M) \ge 1$] at some particular values of $\bp$
due to uncontrolled IR divergences, while remaining good at other values of
$\bp$.  A complete
$\O(g^4)$ calculation allows us to explore larger values of
$T/\bp$, while a partial calculation serves as an indicator of the
reliability of our expansion.

Progress towards our goal is achieved by modifying the propagators $D$, as
in Fig.~2.  The largest $\O(g^3)$ corrections are mass renormalizations,
and are incorporated by replacing $\bm$ in \puteq{vprime} with plasma
masses \cite{fend}.  Working to $\O(g^4)$ requires a self-consistent
solution of the mass gap equations.  These are not the whole story,
however, since we calculated the plasma masses at vanishing external
momenta.  Wavefunction renormalization and the momentum dependence of the
plasma masses must also be included. We plan to address these non-static
corrections in a future publication \cite{bbsv}, but make all effective
potential calculations
in this paper with self energies evaluated at $ \omega_n =0 $ and
three-momenta ${\bf k} =0$.

To put the various corrections in perspective, we estimate their
contributions to the effective potential terms D,E,and $\lambda_T$, as well
as to a ``pseudo-linear'' term J, valid only when the tree mass is larger
than the static plasma mass.  Since daisy and super-daisy contributions are
summed by mass renormalizations (valid at the tadpole level in the
imaginary time formalism, where propagators go like $(k^2 + m^2)^{-1}$), we
schematically insert the relevant mass (see eq. 4.3) into \puteq{vprime},
expand $F({\bm \over T})$ using Appendix B, and read off the J,D,E, and
$\lambda_T$ terms (see Table~1).  The terms involving $\rho$ alone come
from daisies, while those involving $\sigma$ come from super-daisies (and
sub-leading terms in daisies).

\begin{table} \centering \def\g#1{$g^#1$}
\begin{tabular}{|r|cccc|} \hline & $J$ & $D$ & $E$ & $\lambda_T$ \\\hline
                          1-loop &  0  & \g2 & \g3 & \g4         \\
      Daisy ($\rho T^2 < \bp^2$) & \g3 & \g4 & \g7 & \g6         \\
      Daisy ($\rho T^2 > \bp^2$) &  0  & \g3 & \g3 & \g3*        \\
Super-Daisy ($\rho T^2 < \bp^2$) & \g5 & \g4 & \g5 & \g8         \\
Super-Daisy ($\rho T^2 > \bp^2$) &  0  & \g5 &  0  & \g8         \\
                        Lollipop &  0  & \g4 & \g5 & \g6         \\\
$\om \ne 0$ ($\rho T^2 < \bp^2$) & \g3 & \g4 &  0  & \g6         \\
$\om \ne 0$ ($\rho T^2 > \bp^2$) &  0  & \g3 &  0  & \g3*        \\
       $\om = 0, {\bf k} \ne 0$  &  0  & \g4 & \g5 & \g6         \\\hline
  \multicolumn{5}{c}{*Suppressed by powers of $\rho$}            \\
\end{tabular} \caption{Order of Contributions to the Potential}
\end{table}

The largest correction comes from daisies when $\rho T^2 >> \bp^2$,
because the introduction of an infrared cutoff eliminates the cubic term
from the one loop graph.
The plasma mass of the W longitudinal mode corresponds to this case, while
the Higgs plasma mass falls between the two extremes.

All of this is for static self energies.  We can use expressions, valid
in unbroken theories, for the self energy of gauge bosons \cite{weldon}
to estimate the contribution of nonzero frequency $\omega_n \ne 0$, or
nonzero three-momentum ${\bf k} \ne 0$
propagator corrections. A typical correction is a self energy term
$g^2 \rho\, \om^2 {T^2 / {\bf k}^2}$. This gives a contribution
    \Eqa V' &\sim& g^2 \bp \sik [\om^2 (1 + {g^2 \rho T^2 \over {\bf k}^2})
          -{\bf k}^2 - m^2 ]^{-1} \nonumber \\ &\sim&
        g^2 \bp \int_m^T {d^3 k \over \omega_0} F_{\omega_0} \nonumber \\
        &\sim& g^2 T^2 \bp + \rho g^3 \bp T^3 (\rho T^2 + \bp^2)^{-1 / 2}
        \Enda
where $\omega_0^2 = (k^2 + m^2)/(1 + g^2 T^2 {\rho \over k^2})$, and for
purposes of power counting, $F_x$ (defined in Appendix C) $\sim {T / x}$.
For the table, we have assumed an expansion in $m \over T$. In general, the
non-static part of a tadpole is as important as the static part. However,
in the unbroken theory with $\om =0$ (only $\om =0$ contributes to E), the
coefficient of ${\bf k}^2$ is down by $\sim {1 \over \pi^2}$ relative to
the coefficient of $T^2$, suggesting that ${\bp_+ / T_c}$ is not
significantly altered by non-static terms. This contrasts with some recent
claims \cite{evans}.

\begin{figure} 
\vskip2.7in
\caption{Improved Propagators}
\end{figure}

\section{The gauge sector}

To first approximation $\M = m_W = g \bp/2$, so without propagator
modification the loop expansion fails for $\bp < T$, which is unacceptable.
We solve the gap equation of Fig.~2a to obtain both electric and magnetic
plasma masses for the W.  It is known \cite{ftQCD} that for $\bp=0$, to
leading order the magnetic plasma mass vanishes, so we write
  \Eq \Pi_0^0 = \rho_0 \, g^2 T^2 + \sigma_0 \, g^2 m_W T + {\cal O}(g^4
     T^2), \quad \Pi_i^i = \sigma_i \, g^2 T m_W + {\cal O} (g^4 T^2).
  \Endl{rhosig}
and solve
  \Eq m_0 = \sqrt{m_W^2 + \Pi_0^0}, \quad m_i = \sqrt{m_W^2 + \Pi_i^i} \End
The terms neglected (such as contributions of $m_i$ or $m_h$ to $\Pi_0^0$)
are suppressed by powers of $g$,$\lambda$, or $(\M/T)$.  This procedure
yields:

  \Eq m_0 = \frac12 \left[ \sigma_0 g^2 T + \sqrt{(4 \rho_0 g^2 +
     \sigma_0^2 g^4) T^2 + g^2 \bp^2} \right], \Endl{self0}
  \Eq m_i = \frac12 \left[ \sigma_i g^2 T + \sqrt{\sigma_i^2 g^4 T^2 + g^2
     \bp^2} \right]  \Endl{selfi}
We calculate $\rho_0$, $\sigma_0$, and $\sigma_i$ to 1-loop for the Standard
Model in Appendix A, but we will only use the result for $\sin\theta_w =
\lambda = 0$:
  \Eq \rho_0 = 11/6, \qquad \sigma_0 = -5/(4\pi), \qquad \sigma_i = 1/(6\pi)
    \Endl{rss}
We will treat the Z as a third W boson, but implicitly replace $g\to G$.

\begin{figure} 
\vskip2.0in
\caption{Linear Term Tadpole, and Overcounting}
\end{figure}

In ref.~\cite{us} two of the authors used these masses in the
vacuum-to-vacuum W loop, {\it i.e.} in \puteq{Idef}, and found a linear
term in $V$. At the tadpole level, this is equivalent to using both an
improved propagator and an improved three point coupling (Fig.~3a).  For
example, if $\bm^2$ is the plasma mass to $\O(g^4)$, the vacuum to vacuum W
loop generates a tadpole with improved Higgs-W-W coupling $i g_{\mu\nu}
{d\bm^2 \over d\bp}$ and improved inverse propagator $p^2 + \bm^2$. This
overcounts the ``figure eight'' tadpole as shown in Fig.~3b, and is the
source of the linear term (since the improved coupling does not vanish at
$\bp=0$).  It is therefore important to make $\bp$
dependent mass renormalizations at the tadpole level \cite{comm,lpriv}.

It has been suggested \cite{arny,evans} that including the momentum
dependence of the plasma masses eliminates the linear term. While it is
true that the momentum dependence can be an important correction, any $\bp$
dependent mass renormalization, whether or not it also depends on momentum,
must be made at the tadpole (or mass, 3-point, etc.) level to avoid
overcounting.  We believe the linear term found by Shaposhnikov \cite{shap}
arose from a similar overcounting.

Substituting the improved masses into the propagator in
the W tadpole, counting all 3 W's, yields
  \Eq V' = {-3\over2}{g^2 \bp\over2} T \sum_n \int {\d^3 {\bf k} \over
    (2\pi)^3} \Tr\{ P (1-\Pi P)^{-1} \} \End
where $-i P_\mu^\nu (k)$ is the Landau gauge tree-level propagator, and
$\Pi_\mu^\nu = {\rm diag}(\Pi_0^0, \Pi_i^i, \Pi_i^i, \Pi_i^i)$.  This gives
  \Eq V' = {3g^2 \bp\over4} T \sum_n \int {\d^3 {\bf k} \over (2\pi)^3}
    \left[ {3\over k_i^2} - {{\bf k}^2 \delta \over k_i^2 (k_n^2 k_i^2 +
    {\bf k}^2 \delta)} \right] \Endl{vgp}
where $k_n^2 \equiv {\bf k}^2 + (2\pi n T)^2$, $k_i^2 = k_n^2 + m_i^2$,
and $\delta \equiv \Pi_0^0 - \Pi_i^i = m_0^2-m_i^2$.  The first term in
brackets exactly reproduces \puteq{vprime}\ with $\bm\to m_i$.  The
second term, up to renormalizable divergences, is dominated by $n=0$ and is
approximately
  \Eq {3g^2 \bp\over4} \, {T (m_i-m_0) \over 4\pi} \Endl{cubic}
Insofar as $m_i \sim \bp$ and $m_0 \sim T$, this term reduces the cubic term
in $V$ by 1/3, as expected \cite{comm}.  In Appendix C we show there is an
additional (relatively unimportant) correction:
  \Eq {3g^2 \bp\over4} \, {3\delta \over 64\pi^2} \left[ \ln(T^2/M_w^2) +
    5.21 \right] \End

We next turn to the lollipop diagram (Fig.~1e), with propagators improved
to $\O(g^2)$. The contribution of internal $W^{\pm}$'s  to $V'$ is
  \Eqa V'_l &=& {-g^3 M_w \over 2} \sik \sip {1 \over [(p_m -k_n)^2 + m_h^2]}
    \nonumber  \\ && \Tr\{ P(k)[1 - \Pi P(k)]^{-1} P(p)[1 - \Pi P(p)]^{-1} \}
       \Endla{lolli}
There is also a contribution due to the Z. Details of the computation
are left to Appendix C, where the expression is evaluated for static
self energies and $\delta = 0, \infty$.

Finally, we remark on our gauge fixing. In $R_\xi$ renormalizable gauges,
the gauge - Goldstone boson mixing can be eliminated only if one chooses
a different gauge for every value of $\bp$.  Although the potential
can be modified to account for this \cite{dj2}, it is unnecessary in
the case of Landau gauge, where the mixing term vanishes due
to $\partial_\mu A^\mu = 0$.

\section{The Higgs sector}

It is easy to include the effects of gauge bosons and fermion loops on the
Higgs propagator, since the effective potential is the generating
functional of 1PI graphs at zero external momentum.  If $V_G$ is the
potential calculated from gauge bosons and fermions only, then the shifted
Higgs mass $m_h^2 = V_G''$, and the Goldstone boson mass $m_\chi^2 =
V_G'/\bp$.  We could solve the gap equation (to include superdaisies) as we
did for the gauge sector, giving:
  \Eqa m_h^2(\bp,T) &=& V_G''(\bp,T) + {\lambda T^2 \over 4} \left[
    F_+ \left( m_h \over T \right) + F_+ \left( m_\chi \over T \right)
    \right] \nonumber\\
    &+& {\lambda^2 \bp^2 T \over 4} \left[ {3 \over m_h} F'_+ \left( m_h
    \over T \right) + {1 \over m_\chi} F'_+ \left( m_\chi \over T \right)
    \right] \nonumber\\
  m_\chi^2(\bp,T) &=& V_G'(\bp,T)/\bp + {\lambda T^2 \over 12} \left[
    F_+ \left( m_h \over T \right) + 5F_+ \left( m_\chi \over T \right)
    \right] \nonumber\\
    &+& {\lambda^2 \bp^2 T^2 \over 3 (m_h^2-m_\chi^2)} \left[ F_+ \left( m_h
    \over T \right) - F_+ \left( m_\chi \over T \right) \right]
 \Enda
However, except for very heavy Higgses these are well approximated by
  \Eq m_h^2(\bp,T) = V_G''(\bp,T) + {\lambda T^2 \over 2}, \qquad
      m_\chi^2(\bp,T) = V_G'(\bp,T)/\bp + {\lambda T^2 \over 2}
  \Endl{hgapapp}
which corresponds to Fig.~2b.  These are the masses $\bm$ we use in
\puteq{vprime}.  Thus, we have summed contributions to the Higgs mass due
to gauge boson superdaisies, but only Higgs daisies.

Carrington \cite{carr}, working to lowest order in $g$, essentially found
\puteq{hgapapp} but with $V_G''$ and $V_G'/\bp$ replaced by their values at
the origin.  Thus at all interesting temperatures the scalar masses
appeared real.  Our calculation re-introduces imaginary masses; see the
discussion following \puteq{Idef}.  Some bumpiness results in our plots
where $m_h$ and $m_\chi$ pass through zero.

\section{Improving the Standard Model effective potential}

We first calculate the ``1-loop'' effective potential.  Then we omit the
Higgs sector and improve the gauge sector as described in \puteq{vgp}\ and
the subsequent paragraph to get $V_G$.  The Higgs sector is then added back
in using \puteq{hgapapp}\ in \puteq{vprime}, adding the lollipop from
Appendix C, and integrating to get the ``Super-Daisy'' potential, as in
Fig.~2c.

For comparison we also show a ``Daisy'' potential, which differs only in
setting $\sigma_0=\sigma_i=0$ in \puteq{self0}\ and \puteq{selfi} and
omitting the lollipop.  A good ``Estimate'' is obtained by calculating the
1-loop potential with $g_W=4$, $g_Z=2$, and $m_h=m_\chi=0$.

Fig.~4 shows these potentials for various values of the Higgs and top
masses (in GeV), including one set matching Carrington's plots \cite{carr}.
Each potential is shown at its critical temperature $T_1$ (i.e. when two
vacua are degenerate).  $V$, $\bp$, and $T$ are given in units where $v=1$.

For $\bp \gg 2T$ the plasma mass corrections are small, and the 1-loop
potential is adequate.  The perturbative expansion is still out of control
for $g^2 T/m_i > 4\pi$ (where the numerical factor is something of a
guess), or roughly $\bp < .06\,T$ (marked by an arrow on the plots), so
even the ``Super-Daisy'' potential is not to be trusted far to the left of
the arrow.  This is about a factor of 2 closer to the origin than the
corresponding cutoff for the ``Daisy'' potential, $\bp < .10\,T$
\cite{carr}.  Indeed we see that for $M_h > 75\gev$ the ``Daisy'' and
``Super-Daisy'' potentials differ significantly.

We plot $\phi_+(T_1)/T_1$ (which closely approximates $\phi_+(T_b)/T_b$)
vs. $M_h$ for several values of $M_t$ in Fig.~5.  For comparison we show
values from both Carrington \cite{carr} Fig.~14 (with $M_t=110\gev$), and
Dine {\it et al.} Fig.~5 ($M_t=120\gev$); in the latter case we converted
their results at $T_b$ (which they call $T_t$ for ``tunneling'') using the
quartic potential relation
  \Eq {\phi_+(T_1) \over T_1} = {\phi_+(T_b) \over T_b} \left[ 4 \over
    3 + \sqrt{1+8\epsilon_b} \right], \qquad
  \epsilon_b = {T_1^2 - T_b^2 \over T_1^2 - T_2^2} \End
and took $\epsilon_b$ from their Fig.~6.  For a light Higgs, where
$\O(g^4)$ corrections are small (see Fig.~4), all the results agree
closely.  As the Higgs mass increases, higher-order corrections appear to
lower $\phi_+(T_1)/T_1$.  We note the top mass is nearly irrelevant for
heavier Higgses.  Since $\phi_+(T_1)/T_1 \ll 1.4$ for all
experimentally-allowed Higgs and top masses, we see the EWPT remains too
weakly first-order to drive baryogenesis.

\begin{figure} \vskip\vsize \end{figure} 
\begin{figure} 
\vskip6.8in
\caption{1-Loop, Daisy, Super-Daisy, \& Estimate Potentials}
\end{figure}

\begin{figure} 
\vskip2.9in
\caption{$\phi_+/T_1$ vs. $M_h$}
\end{figure}

\section{Saddlepoints and the effective action}

While the effective potential suffices to determine the order of the
transition, the full action $\Gamma$ [see \puteq{Gamma}] is needed to
determine the dynamical properties of the system, such as the rates for
bubble nucleation or sphaleron fluctuations.  For extremal configurations
such as critical bubbles, $J=0$ in \puteq{Legend} and $\Gamma[\bp] = W[0]$
is gauge invariant \cite{ginv}, so derivative terms must cancel out the
gauge dependence of $V$.

We can use Chan's derivative expansion \cite{chan} to estimate the size of
derivative corrections.  Since the finite temperature Green's functions
$G(x,y)$ satisfy the same equations as the zero temperature ones, but with
periodic boundary conditions, we may formally express $G(x,y)$, in Landau
gauge, as
\Eq G(x,y) = T \sum_n \int \frac{d^3 p}{(2 \pi)^3} e^{i p \cdot y}
    [ -\partial_x^2 + U(x) ]^{-1} e^{- i p \cdot x}, \End
where $U(x)$ is the mass of the field in question, including $\bp$
independent plasma corrections.  Following Chan's technique, we write
$G(x,x)$ as an explicit expansion in even powers of derivatives:
\Eqa G(x,x) &=& T \sum_n \int \frac{d^3 k}{(2 \pi)^3}
   [ k^2 + U(x + i \frac{\partial}{\partial k} ) ]^{-1} \nonumber\\
       &=& T \sum_n \int \frac{d^3 k}{(2 \pi)^3}
   [ k^2 + U(x)]^{-1} \sum_m^{\infty} \Bigl( - \sum_q^{\infty}
    \frac{1}{q!} (\partial_{j_1}...\partial_{j_q} U(x)) \nonumber\\
  & & \left[  i \frac{\partial}{\partial k_{j_1}}...i
     \frac{\partial}{\partial k_{j_q} }   \right]
  [ p^2 + U(x)]^{-1}  \Bigr)^m.
\Enda

The contribution to the spatial part of the kinetic energy
arises from the $m=1, q=2$ and $m=2, q=1$ terms in the above sum.
Taking only the dominant $n=0$ part of the frequency sum gives
\Eqa {\cal L}_{eff} ~&\ni&~ \int \delta U(x) G(x,x) \nonumber \\
                &=& {-g_j T \over 384 \pi} U^{-3/2} (U')^2
     (\partial_i\bp)^2 = {-1 \over 64\pi}{g^2 T \over (g\bp/2)}
     (\partial_i
     \bp)^2
\Endla{dterms}
where $U = m_j^2(\bp)$, and $\partial_i$ is a spatial derivative.  We have
inserted values for the tree level $W$ mass in the second expression,
corresponding to the $\p^2 $ term of the penultimate diagram in Fig.~2b.

The form is as expected by naive power counting of a one
loop graph with two external legs, both carrying nonzero momentum.
Since this $\O(g)$ contribution is numerically small, one might think
derivative corrections are unimportant. That this may {\it not} be
the case is indicated by the calculation of the $\O(g^2)$ graph in
Fig.~6.

\begin{figure} 
\vskip1.0in
\caption{Momentum-dependent Higgs Self-Energy at $\O(g^2)$}
\end{figure}

This graph arises in the derivative expansion from a shifting
of $G^{-1}$ due to mixing which we previously ignored.
Its contribution (setting $s_w =0$) to the real part of the Higgs
self-energy, with real external four momentum $p$, is
\Eqa \Pi(p_0,{\bf p})  &=& 3 g^2 \sik {p^2 - {(p \cdot k)^2 \over
k^2}
  \over  (k^2 + m_W^2)[(p+k)^2 + m_{\chi}^2] } \nonumber \\
  &=& 3 g^2 \{ [\frac58 p^2 + \frac{\mc - \mw}8 ] L(\frac{\mw -\mc
-p^2}2;
   p_0,\p) \nonumber \\ &+& [\frac{\mc}8 - \frac38 p^2]
   L(\frac{-p^2 - \mc}2;p_0,\p)
   \nonumber \\ &-& [{(p^2 -\mc)^2 \over 8 \mw}] L(\frac{p^2 -
\mc}2;p_0,\p)
   \nonumber \\ &+& [\frac{(p^2 -\mc +\mw)^2}{8 \mw} - \frac12 p^2]
   L(\frac{p^2 +\mw - \mc}2,p_0,\p) \nonumber \\ &+&
   [{ p^4 + \mc (\mc -\mw) + p^2 (3 \mw - 2 \mc) \over 16 (
   p^2 + \mc - \frac{\mw}2 ) }] [ L( \frac{\mw}2 -p^2 - \mc;p_0,\p)
   \nonumber \\ &+& L(0;-p_0,\p) ] \} +
   \O(\frac{m^2}{T^2})
     \Enda
in which
\Eq L(m^2;p_0,\p) = \int_0^\infty {d k \over 2 \p (2 \pi)^2}
      [F_k \ln|{k (\p - p_0) - m^2 \over k (\p +p_0) + m^2} | +
      F_{-k} \ln|{k (\p - p_0) + m^2 \over k (\p +p_0) - m^2} | ]
      \End
and $\p$ is the magnitude of the the external three-momentum.

The integral has been computed previously \cite{weldon}, to
$\O(T^{-2})$. After a minor algebraic correction, it gives
 \Eqa  L(m^2;p_0,\p) &=& {T \over 8 \p} \theta(-p^2) \,{\rm sign}(-m^2)
          \nonumber \\
          &+& {m^2 \over 4 \pi^2 p^2} [{p_0 \over 2 \p}
            \ln |{p_0 + \p \over p_0 - \p}|
          + \gamma_{\rm Euler} -1
          + \frac12 \ln {M_W^2 \over 4 \pi^2 T^2 }]
       \Enda
where $M_W$ is our subtraction point, and ${\rm sign}(0)=0$.  The analytic
behavior of this diagram [note the strange $T / \p$ behavior of
$L(m^2;p_0,\p)$] is sensitive to the scheme one uses to continue from
imaginary to real external momentum.  Our continuation prescription is
consistent with previous work \cite{weldon,wel2}, but there are other
methods \cite{analy}. The $p_0 =0, \p \ne 0$ behavior, which characterizes
saddlepoint solutions, may be altered if our prescription turns out to be
incorrect.

\section{Conclusion}

Much recent work on the EWPT \cite{us,shap,carr,comm,evans} has
concentrated on improving the calculation of V. We now believe there is
no linear term, and that the propagator improvement performed in
ref.~\cite{carr} and estimated in ref.~\cite{comm} is essentially correct
for the gauge sector to $\O(g^3)$.  The main result of these corrections is
to screen the longitudinal mode, decreasing the cubic term $E$ by a factor
of $1/3$ and making the transition more weakly first-order.

In this paper we have included higher order corrections not previously
considered, specifically those from subleading parts of daisy graphs,
gauge superdaisies, gauge superdaisies in the Higgs sector,
and the ``lollipop'' diagram.  We estimated the effect on the effective
potential of using momentum-dependent self-energies,
and computed the $\O(g^2)$ derivative corrections to the effective action.
The results of our effective potential computations are similar to those of
\cite{carr,comm} for a light Higgs ($M_h \le 75\gev$), but show an even
further weakening of the transition for a heavier Higgs ($75\gev < M_h
< 125\gev$).  Above $125\gev$ our expansion becomes less reliable.

Evans \cite{evans} has criticized all recent calculations of the
electroweak effective potential on the grounds that propagator resummations
have been performed at zero external momentum, rather than on-shell. We
find that although this approximation does lead to errors (as pointed
out in our earlier preprint \cite{us}), they are unlikely
to lead to any qualitative changes in the behavior of the potential.

Derivative corrections to the effective action may be important, however.
For example, if the diagram in Fig.~6 is indicative, they could
significantly alter bounce solutions when
typical spatial frequencies are less than about $T / 8$. Estimates of
bubble wall thicknesses are often tens of $T^{-1}$ \cite{mcl,comm}.
Similar corrections to the gauge boson effective action could
distort the sphaleron solution.
We hope to examine derivative corrections and their consequences
in more detail in the future \cite{bbsv}.

To summarize, a computation of $V$ in any finite loop approximation is
subject to uncontrolled infrared corrections for $\bp < T$.  When plasma
masses are included in both the gauge and Higgs sectors, the improved
potential is reliable for $\bp > gT/10$.
We have computed $V$ for the Standard Model, and compared the
results to previous estimates. If current estimates of sphaleron energies
are reliable, the Standard Model (even with augmented CP
violation) is still inadequate to generate the baryon asymmetry.

\vskip 1.cm
\section{Acknowledgments}

The authors would like to thank Greg Anderson, Peter Arnold, Meg Carrington,
Michael Dine, Thomas Gould, Lawrence Hall, Clarence Lee, Andrei Linde,
Ann Nelson, Stamatis Vokos, Erick Weinberg, and Mark Wise for numerous
discussions.
DEB acknowledges support from the U.S.  Department of Energy under Contract
DEAC-03-81ER40050, and thanks the Stanford Linear Accelerator Center
for its hospitality during part of this work.
SDH acknowledges support from the National Science Foundation under grant
NSF-PHY-87-14654, the state of Texas under grant TNRLC-RGFY106, and from
the Harvard Society of Fellows.
CGB acknowledges support from the U.S. Department of Energy under Contract
DEFG-02-90ER40560, as well as the National Science Foundation under
grant NSF-PHY-91-23780.

\newpage
\section{Appendix A:  Standard Model $f_j$'s and $\Pi$'s}

The functions $f_j(r)$ defined in \puteq{fjdef}\ are (to 1-loop)
  \Eqa f_h(1) &=& 18 (\pi/\sqrt3 - 2) \nonumber\\
    f_\chi(r) &=& {8 \lambda^2 v^4 \over M_\chi^4} \left[ \ln(r) + i\pi - 2
      \right] \nonumber\\
    f_t(r) &=& {4(4-r)^2 \over \sqrt{4r-r^2}} \tan^{-1} \left( r \over
      \sqrt{4r-r^2} \right) - 16 + 2r \nonumber\\
    f_W(r) = f_Z(r) &=& {4\over3} \sqrt{4-r \over r} (r^2-4r+12) \tan^{-1}
      \left( r \over \sqrt{4r-r^2} \right) + \nonumber\\
      && {4 (1-r)^3 \over 3r} \ln(1-r) - {44 \over 3} + {2r^2 \over 3}
      (\ln(r) - i\pi)  \Enda
Note the imaginary part of $f_\chi$ (representing the amplitude for a Higgs
to decay to Goldstone bosons in the ungauged theory) is exactly canceled
by a term in $f_{W,Z}$.  For large $r$, the leading results for the top and
the gauge bosons (corresponding to taking $\Sigma(M_h^2)-\Sigma(0) = M_h^2
\Sigma'(0)$) are
  \Eq f_t(r) \approx -10r/3, \quad f_W(r)=f_Z(r) \approx -10r/3 \End

The electric and magnetic plasma masses for the $W^\pm$ from Fig.~2a are:
  \Eqa \Pi_0^0 (W^\pm) &=&
    {g^2 T^2 \over 3} \left[ 2 + {1\over2} + {12\over4} \right]
  - {g^2 T \over 2\pi} \left[ m_W + c_w^2 m_Z + s_w^2 m_\gamma +
      {m_h + 3m_\chi \over 8} \right] \nonumber\\
  &-& {g^4 \phi^2 T \over 16\pi} \left[ {1\over m_h + m_W} + {s_w^2 \over
    m_\chi + m_\gamma} + {s_w^4/c_w^2 \over m_\chi + m_Z} \right]
  \nonumber\\
  \Pi_i^i (W^\pm) &=& {g^2 T \over \pi}
    \Bigl[ {-7\over12}(m_W + c_w^2 m_Z + s_w^2 m_\gamma)
   + c_w^2 {m_Z^3-m_W^3 \over m_Z^2-m_W^2} + s_w^2 {m_W^3-m_\gamma^3 \over
     m_W^2 - m_\gamma^2} \nonumber\\
  & &\quad - {m_h + m_\chi \over 16} + {1\over12}{m_h^3-m_\chi^3 \over
   m_h^2-m_\chi^2} \Bigr] \nonumber\\
  &-& {g^4 \phi^2 T \over 24\pi} \left[ {1\over m_h + m_W} + {s_w^2 \over
    m_\chi + m_\gamma} + {s_w^4/c_w^2 \over m_\chi + m_Z} \right]  \Enda

The three numbers in brackets in the first term of $\Pi_0^0$ reflect
contributions from the gauge sector, the Higgs sector, and 12 fermionic
isospin doublets, respectively.

The analogous formulas for the $Z^0$ are
  \Eqa \Pi_0^0 (Z^0) &=&
    {g^2 T^2 \over 3} \left[ 2 c_w^2 + {1\over2}\; {1 - 2 s_w^2 c_w^2 \over
      c_w^2} + {12\over4}\; {1 - 2 s_w^2 + 4 s_w^4 \over c_w^2} \right]
      \nonumber\\
  &-& {g^2 T \over 2\pi} \left[ 2 c_w^2 m_W + {m_h + (3 - 8 s_w^2 c_w^2)
      m_\chi \over 8}  \right] \nonumber\\
  &-& {g^4 \phi^2 T \over 16\pi} \left[ {1/c_w^4 \over m_h + m_Z} + {2
    s_w^4/c_w^2 \over m_\chi + m_W} \right] \nonumber\\
  \Pi_i^i (Z^0) &=& {g^2 T \over \pi} \left[ {c_w^2 \over 3} \, m_W
    - {m_h + m_\chi \over 16 c_w^2} + {1\over12 c_w^2} \;
      {m_h^3-m_\chi^3 \over m_h^2-m_\chi^2} \right] \nonumber\\
  &-& {g^4 \phi^2 T \over 24\pi} \left[ {1/c_w^4 \over m_h + m_Z} + {2
    s_w^4/c_w^2 \over m_\chi + m_W} \right] \Enda
and for the photon
  \Eqa \Pi_0^0 (\gamma) &=& {e^2 T^2 \over 3} [2 + 1 + 12]
    - {e^2 T \over 2\pi} \left[ 2 m_W + m_\chi \right]
    - {e^2 g^2 \phi^2 T \over 8\pi (m_\chi + m_W)} \nonumber\\
  \Pi_i^i (\gamma) &=& {e^2 T \over \pi} \left[ {1\over3} m_W \right]
    - {e^2 g^2 \phi^2 T \over 12\pi (m_\chi + m_W)} \Enda

To simplify things, we take $s_w\to0$ (with $g$ constant) and
$\lambda\to0$.  The former approximation introduces an error of $\O (e) =
\O (g s_w)$ in the plasma mass solutions, which we neglect.  Then the
$W^\pm$ and $Z^0$ self energies both become
  \Eq \Pi_0^0 \approx {11 g^2 T^2\over6} - {5g^2 T\over4\pi} m_W, \qquad
      \Pi_i^i \approx {g^2 T\over 6\pi} m_W \End
which gives \puteq{rss}.

\newpage
\section{Appendix B: Numerical approximations for $I_\pm$}

The following approximations \cite{dj,AH,greg,haber} to $I_\pm$ [see
\puteq{Idef} and the following paragraph] are accurate to $10^{-4}$:
  \Eq I_+(y<1) \approx {-\pi^4\over45} + {\pi^2\over12}y^2 - {\pi\over6}y^3
    - {y^4\over32} (\ln y^2 - 5.4076) + .00031 y^6 \End
  \Eq I_-(y<1) \approx {-7\pi^4\over360} + {\pi^2\over24}y^2
    + {y^4\over32} (\ln y^2 - 2.6350) - .00214 y^6 \End
  \Eqa I_+(i(y<1)) &\approx& {-\pi^4\over45} - {\pi^2\over12}y^2 + y^3 \left[
    {4\over9} - {1\over3} \ln(2y) \right] - {y^4\over32} (\ln y^2 - 5.4076)
    \nonumber\\
    & &+ {y^5\over180} - .00029 y^6 \Enda
  \Eq I_+(y>1) = -\sum_{n=1}^\infty {y^2 K_2(ny) \over n^2} \End
  \Eq I_-(y>1) = \sum_{n=1}^\infty (-1)^n {y^2 K_2(ny) \over n^2} \End
  \Eqa && I_+(i(y>1)) = \sum_{n=1}^{[8/y]} \left\{ {y^3\over3n} -
      {\pi y^2\over2n^2} \left[ {\bf H}_2(ny) - Y_2(ny) \right] \right\}
      \nonumber\\
   && \,\, + {y\over\pi} \sum_{j=0}^4 \left\{ \left[ \zeta(2j+3) -
      \sum_{n=1}^{[8/y]} n^{-(2j+3)} \right] \left[ -4\over y^2 \right]^j
      \Gamma(j+\frac32) \Gamma(j-\frac12) \right\} \Enda
where $K$ and $Y$ are Bessel functions, $\bf H$ is the Struve function,
$\zeta$ is the Riemann zeta function, $[x]$ is the greatest integer less
than or equal to $x$, and infinite sums are terminated when the desired
accuracy is achieved.

We note that the $\phi^4 \ln \phi^2$ terms which come from $I_\pm(g\phi/T)$
and those which come from the $T\!=\!0$ potential cancel.

\newpage
\section{Appendix C: Gauge tadpole and lollipop}

We may rewrite \puteq{vgp}\ as
  \Eq V' = {3g^2 \bp\over4} T \sum_n \int {\d^3 {\bf k} \over (2\pi)^3}
    \left[ {2\over k_i^2} + {k_n^2 \over k_n^2 k_i^2 + {\bf k}^2 \delta}
    \right] \Endl{vgp2}
and evaluate the sum with the usual contour integral trick. The second
term then becomes
  \Eq  {3g^2 \bp\over4} \int {\d^3 {k} \over (2\pi)^3}
     \left[ {k_+ + k_- + {m_i^2 \over \Delta}(k_- - k_+) \over
     4 \sqrt{k^4 + m_0^2 k^2} }  +
     {m_i^2 \over 2 \Delta} ({F_{k_+} \over k_+} - {F_{k_-} \over k_-})
     + {F_{k_+} \over 2 k_+} + {F_{k_-} \over k_-} \right]
     \Endl{vgp3}
where $\Delta = \sqrt{m_i^4 - 4 \delta k^2},$
$k_{\pm} = \sqrt{ k^2 + {m_i^2 \over 2} \pm {\Delta \over 2} },$
and $F_x = [ e^{x / T} -1 ]^{-1}$.

After renormalization, \puteq{vgp3}\ is well approximated by
   \Eq {m_i^2 \over 16 \pi^2}[\ln{m_i^2 \over M_w^2} - .39] +
     {3 \delta \over 64 \pi^2}[\ln{T^2 \over M_w^2} + 5.21]
     + {T^2 \over 12} F({m_i \over T}) + {T \over 4 \pi}(m_i - m_0)
     \Endl{vgp4}
Note that although the $k_{\pm}$ can be complex, the final answer is real.

The lollipop, \puteq{lolli}\ , has been evaluated elsewhere
in unitary gauge \cite{as} with unimproved propagators.  In Landau gauge,
it is somewhat more complicated.  For static self energies,
 \Eq \Tr\{ P(k)[1 - \Pi P(k)]^{-1} P(p)[1 - \Pi P(p)]^{-1} \}=
   {1 \over k_i^2  p_i^2} [2 + {(k_n \cdot p_m)^2 \over k_n^2 p_m^2}
   - X \delta] \Endl{loll1}
where $p_i^2 = p_m^2 + m_i^2$ and
 \Eq X = (k_n \cdot p_m)^2 {k_\delta^4 {\bf p}^2 + p_\delta {\bf k}^2
     - {\bf k}^2 {\bf p}^2 \delta \over k_n^2 k_i^2 k_\delta^4
     p_n^2 p_i^2 p_\delta^4 }      +
   ({\bf k}^2{ \bf p}^2 - ({\bf k} \cdot {\bf p})^2) { k_i^2 + p_i^2 + \delta
     \over k_i^2 p_i^2 k_\delta^4 p_\delta^4}  \Endl{loll2}
in which $k_\delta^4 = k_n^2 k_i^2 + {\bf k}^2 \delta$.

The frequency sums can be evaluated by using the contour trick:  For
$ f(p) = {g(p;k)/ [(p-k)^2 -m^2]}$, $p_0 = 2 \pi i n T$, and
$g(p,k)$ non-singular at the explicit pole, the contribution from the pole is
 \Eq \sip f(p) = -\int {d^4 p \over (2 \pi)^3} \,{\rm sign}(p_0)
    \delta(p^2 - m^2) F_{p_0 + k_0} g(p+k;k)   \Endl{contour}
For $k_0 =2\pi imT$ (i.e., an internal momentum), $F_{p_0 + k_0} = F_{p_0}$.
This formula should be applied to each pole in \puteq{loll1}.

Temperature-dependent infinities are canceled by graphs analogous
to the lollipop, but with insertions of zero temperature counterterms.
The finite part of \puteq{lolli} is, for $\delta =0$,
 \Eqa V'_l &=& {g^3 M_w \over 2} \{ (3/2 - {m_h^2 (m_i^2 + m_h^2)
   \over 2 m_i^4}) A({m^2 \over 2})  +  {(m_i^2 -m_h^2)^2 \over
   2 m_i^4} A({m_h^2 -m_i^2 \over 2})\nonumber  \\
   &&- (3/4 + {m_h^4 \over 4 m_i^4}
   ) A({m_h^2 + m_i^2 \over 2})  \nonumber \\ &&+ ({m_h^2 \over 2 m_i^2} -
   {m_h^4 \over 8 m_i^4} - 3/2) A({m_h^2 - 2 m_i^2 \over 2}) \}
   \Enda
where
 \Eqa  A(m) &=& \int_0^{\infty} {dk dp \over (2 \pi)^4} F_{k_0} F_{p_0}
   \ln{|{m^2 - 2 p k \over m^2 + 2 p k}|}    \\
        &\approx & 3.11 \cdot 10^{-5} T^2 (\ln{m^2 \over T^2} -1 )
    \Enda
For $\delta \to \infty$,
  \Eq V'_l = {g^3 M_w \over 2} [B(m_h^2,{m_h^2 \over 2})
    - {1\over 2} B(m_i^2,{m_h^2 - 2 m_i^2 \over 2}) ]  \Endl{B}
where
  \Eqa  B(m^2,M^2) &=& \int_0^{\infty} {dk dp \over (2 \pi)^4} F_a F_b
       {k p \over 2 a b} \{ {4 M^2 \over k p} \nonumber\\ &+&
       [1 + {(M^2 - a b)^2 \over k^2 p^2}] \ln|{k p + M^2 - a b
       \over k p - M^2 + a b}|  \nonumber \\ &+&
       [1 + {(M^2 + a b)^2 \over k^2 p^2}] \ln|{k p + M^2 + a b
       \over k p - M^2 - a b}| \}   \Enda
and $a^2 = k^2 + m_i^2$, $b^2 = p^2 + m^2$.  A numerical approximation
to \puteq{B} is
    \Eq V'_l \approx {g^3 M_w \over 2} .002 T^2 \left[ {m_h \over m_i}
        e^{-.9 (m_h + m_i)/T}  -  {m_h^2 - 2 m_i^2 \over 2 m_i^2}
        e^{-1.8 m_i/T}   \right]    \Endl{lollinf}
We use $\delta =\infty$ in all plots in this paper.


\newpage
\def\ap#1#2#3{           {\it Ann. Phys. }{\bf #1}, #2 (19#3)}
\def\com#1#2#3{          {\it Comm. Math. Phys. }{\bf #1}, #2 (19#3)}
\def\np#1#2#3{           {\it Nucl. Phys. }{\bf #1}, #2 (19#3)}
\def\pl#1#2#3{           {\it Phys. Lett. }{\bf #1}, #2 (19#3)}
\def\pr#1#2#3{           {\it Phys. Rev. }{\bf #1}, #2 (19#3)}
\def\prep#1#2#3{         {\it Phys. Rep. }{\bf #1}, #2 (19#3)}
\def\prl#1#2#3{          {\it Phys. Rev. Lett. }{\bf #1}, #2 (19#3)}
\def\rmp#1#2#3{          {\it Rev. Mod. Phys. }{\bf #1}, #2 (19#3)}
\def\zp#1#2#3{           {\it Z. Phys. }{\bf #1}, #2 (19#3)}

\end{document}